\documentclass[twocolumn,a4paper,aps,prl,showpacs,superscriptaddress,floatfix] {revtex4}

\usepackage{bm}
\usepackage[windows]{umlaute}
\usepackage[dvips]{epsfig}
\usepackage{graphicx}
\usepackage{amsmath}
 
\begin{document}

\title{Role of electric field in surface electron dynamics  above the vacuum level}

\author{J.I. Pascual}
\affiliation{Institut f\"ur Experimentalphysik, Freie
Universitä\"at, Arnimalle 14, D-14195 Berlin, Germany}

\author{C. Corriol}
\affiliation{Donostia International Physics Center (DIPC), San
Sebastian, Spain}

\author{G. Ceballos}
\affiliation{Istituto Nazionale per la Fisica della Materia TASC, Area
Science Park, I-34012 Trieste, Italy}

\author{I. Aldazabal}
\affiliation{Departamento de
F\'{\i}sica de Materiales UPV/EHU, San Sebastian E-20080, Spain}

\author{H.-P. Rust}
\affiliation{Fritz-Haber-Institut der Max-Planck-Gesellschaft, Faradayweg 4-6
D-14195 Berlin, Germany}

\author{K. Horn}
\affiliation{Fritz-Haber-Institut der Max-Planck-Gesellschaft, Faradayweg 4-6
D-14195 Berlin, Germany}

\author{J. M. Pitarke}
\affiliation{Materia Kondentsatuaren Fisika Saila, UPV/EHU, Bilbao, Spain}
\affiliation{Unidad de F\'{\i}sica de Materiales,
Centro Mixto CSIC-UPV/EHU,
San Sebasti\'an, Spain}

\author{P. M. Echenique}
\affiliation{Unidad de F\'{\i}sica de Materiales,
Centro Mixto CSIC-UPV/EHU,
San Sebasti\'an, Spain}
\affiliation{Departamento de F\'{\i}sica de Materiales UPV/EHU,
San Sebastian E-20080, Spain}

\author{A. Arnau}
\affiliation{Unidad de F\'{\i}sica de Materiales,
Centro Mixto CSIC-UPV/EHU,
San Sebasti\'an, Spain}
\affiliation{Departamento de F\'{\i}sica de Materiales UPV/EHU,
San Sebastian E-20080, Spain}

\begin{abstract}

Scanning tunneling spectroscopy (STS) is used to study the dynamics
of hot electrons trapped on a Cu(100) surface in field emission
resonances (FER) above the vacuum level. Differential conductance
maps show isotropic electron interference wave patterns around
defects whenever their energy lies within a surface projected band
gap. Their Fourier analysis reveals a broad wave vector 
distribution, interpreted as due to the lateral acceleration of hot 
electrons in the inhomogeneous tip-induced potential. A line-shape 
analysis of the characteristic constant-current conductance spectra 
permits to establish the relation between apparent width of peaks 
and intrinsic line-width of FERs, as well as the identification of 
the different broadening mechanisms.


\end{abstract}

\maketitle

\section{INTRODUCTION}

A detailed knowledge of the electron dynamics at surfaces is crucial for
an understanding of a large variety of processes, ranging from electron
scattering at surfaces to charge transport dynamics across
interfaces, relevant to design electronic devices \cite{nienhaus,pedro}.
Electrons trapped in unoccupied long-lived resonances represent an interesting
workbench. They favour the localization of photo-injected electrons at
molecular resonances, thus enhancing the catalytic activity of metals
\cite{bovensiepen}. They also represent a valuable probe to investigate
the rich phenomenology behind charge injection and hot-electron quenching.
Experimental techniques such as inverse photoemission \cite{himpsel},
two-photon photoemission \cite{boger}, or ballistic electron scattering
\cite{kasemo} have been traditionally used to study hot electron dynamics
at surfaces.

Recently, scanning tunneling spectroscopy (STS) has proved to be a useful
tool to provide quantitative information about the electronic
structure\cite{crommie,hasegawa,hofmann,petersen, pascualprl} and also the
electron and hole dynamics at metal surfaces \cite{burgi,kliewer,
kern,corriol}. In most cases, these studies have been restricted to low
applied bias voltages, where the applied electric field does not play an
important role \cite{berndt_stark}. A renewed interest has emerged in
using the STM in the field-emission regime, i.e., at bias voltages larger
than the tip work function. In this regime, the applied electric field
lifts up the potential barrier above the vacuum level of the sample,
introducing a new class of resonances that are absent at low bias
voltages, the so-called field-emission resonances (FER)
\cite{binig,becker}. In previous studies, FERs have been used to explore
local changes of the surface work function \cite{himpsel2,schneider},
scattering properties of surfaces and interfaces \cite{kubby,moleetal},
and to achieve atomic-scale imaging of  diamond \cite{bobrov}. A promising
application of FERs is to provide information about the dynamics of
electrons in image states at surfaces \cite{kern}. This is intriguing
since FERs are a characteristic of the tip-induced potential barrier
itself and, therefore, they would exist even in the absence of an image
potential. Hence, a model is needed which describes the dependence of
field emitted electrons dynamics along the surface  on the topology of the
surface potential  and accounts for STS spectra in a wide sample bias
range.

In this paper we demonstrate that electrons trapped in long-lived FERs
are sensitive to the potential gradient induced by the STM tip along the
surface. Scattering of quasi-free FERs electrons with surface defects
gives rise to isotropic two-dimensional (2-D) wave patterns, whose wave
vector components in reciprocal space reflect the local perturbation of
the surface image potential induced by the STM tip. A combined
theoretical and experimental analysis of the FERs peaks in dI/dV
spectra reveals that their line-shape carries information about the
scattering properties of the surface, and hence, about their band
structure. Our calculations permit to identify the different intrinsic and
extrinsic broadening mechanisms of peaks associated with FERs in
conductance spectra in a wide energy range. The organization of the
paper is as follows.  Section \ref{section2} describes the way the experiments were
done. In section \ref{section3} we present the results and discussion of them in
two subsections: subsection \ref{subsection1} is devoted to the analysis of wave
patterns that appear in dI/dV maps, while subsection \ref{subsection2} presents a
line-shape analysis of dI/dV spectra. Finally, in section \ref{section4} the
conclusions of our work are presented.

\section{EXPERIMENT \label{section2}}

The experiments were performed in a custom made ultra-high vacuum scanning
tunneling microscope in thermal equilibrium with a liquid helium bath
\cite{horn}. All spectroscopy data presented in this work were acquired at
4 K. The Cu(100) sample surfaces were cleaned by repetitive cycles of
${Ar}^{+}$ sputtering (1keV) and annealing at 900 K.  The differential
conductance (dI/dV) was measured using a lock-in amplifier above the
low-pass frequency of the feed back loop (f$_{ac}\sim$3 kHz).
The dI/dV spectra shown here are taken in constant current mode
(feed back loop closed) \cite{binig,becker}.

\begin{figure*}[tbp]
\includegraphics*[width=0.5\linewidth] {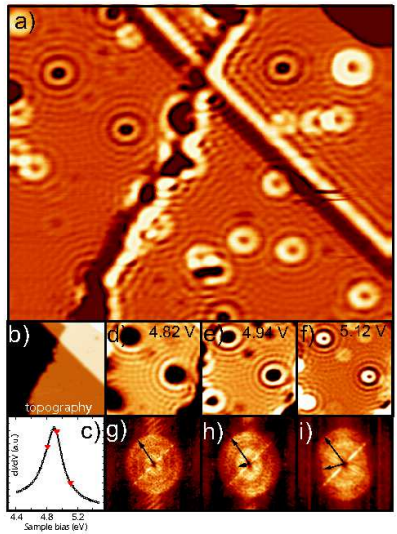}
\caption{(Color online) a) dI/dV map of the Cu(100) surface shown in
b) (I=10 nA; V$_S$=4.9 V). Standing waves are clearly seen around
steps, surface and sub-surface defects. c) Constant current dI/dV
spectra around the first field emission resonance. d-f) dI/dV maps
of a smaller region taken at the bias values indicated in the
figure. g-i) 2-D Fourier transform of d-f). The arrows indicate the
inner and outside radius of the disk appearing in K-space (see the
text). }
\end{figure*}

\section{RESULTS AND DISCUSION \label{section3}}

\subsection {Wave patterns \label{subsection1}}

The electron dynamics of FER states is essentially quasi-free in the
plane parallel to the sample surface, since the corresponding wave
functions lie mainly on the vacuum side of the surface. Therefore,
these electron states, similar to image-states, are not affected by
the corrugation of the surface. As for the case of low lying
surface states, hot electrons in field emission resonances are
expected to have a lifetime long enough to be scattered by steps and
defects at the surface, giving rise to a characteristic standing
wave patterns. In Fig. 1a a constant current dI/dV map shows clear
2-D wave patterns around steps and point defects on the Cu(100)
surface (Fig. 1b). The image is measured with an applied bias
voltage (V$_s$ of 4.9 V, corresponding to the position of the 1st
FER peak (Fig. 1c). A monotonic change to shorter wavelengths with
the applied bias (Figs. 1d-1f) reflects the energy dispersion of
these states. At a first glance the dI/dV maps seem similar
to those taken on the (111) faces of noble metals at lower bias.
However, an important difference becomes apparent when
looking at their two-dimensional (2D) Fourier Transformation (FT)
(Figs. 1g-1i): here the electron wave vector is not constrained to
one single k(E) value but shows a broad distribution, causing that
the 2D FT maps resemble a disk instead of a ring \cite{petersen}.

\begin{figure*} [tbp]
\includegraphics*[width=0.5\linewidth] {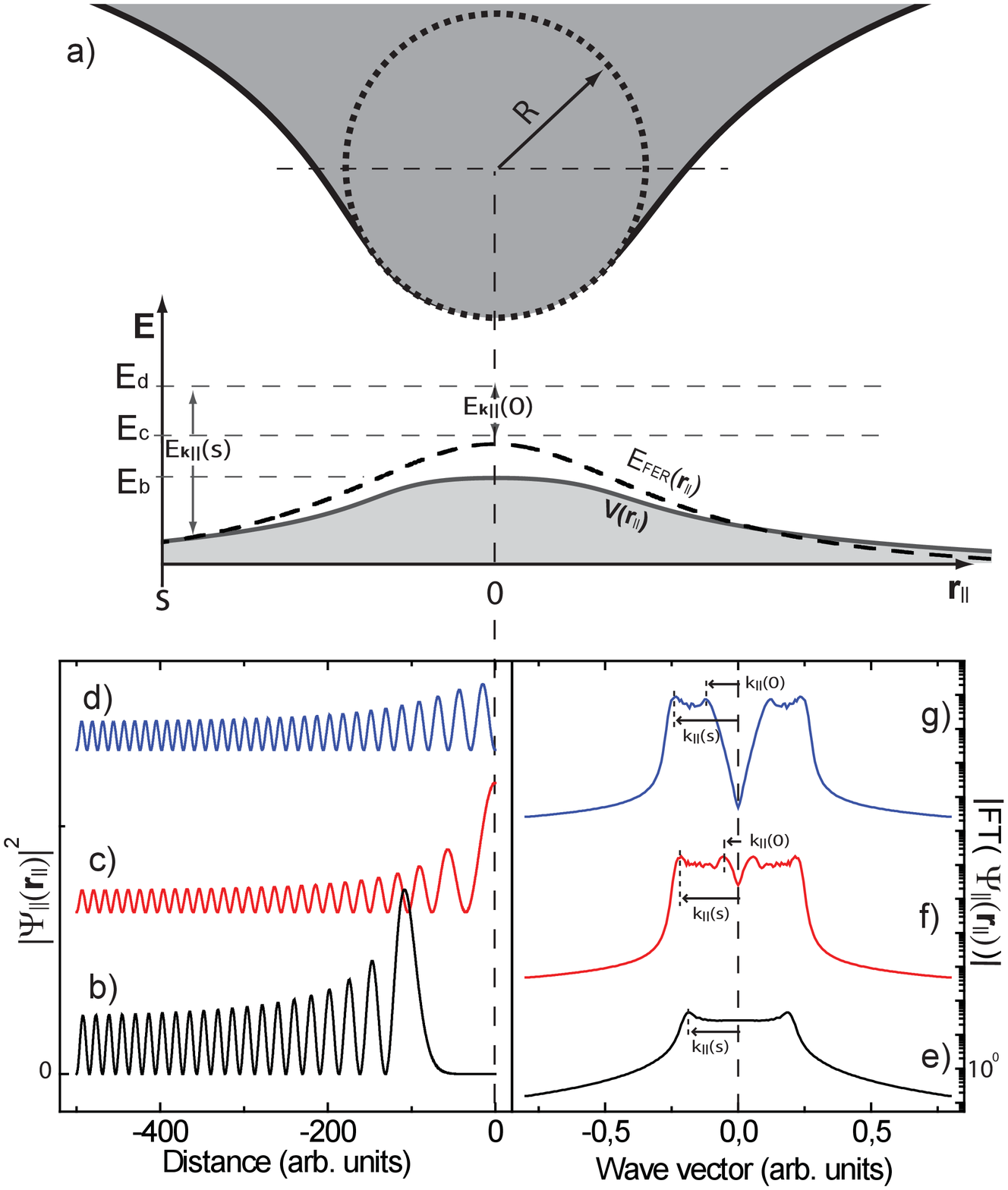}
\caption{(Color online) a) Schematic representation of the
non-conservation of parallel momentum due to the applied field by
a tip of finite radius R. At a given value of electron energy
$E=eV_s-\phi$ the kinetic energy for parallel motion $E_{k_{||}}$
increases from the tip position $r_{||}=0$ to distant sites at the
surface $r_{||}=s$. b), c) and d) show the charge density of electron
states confined in a large box with a potential
$V(x)=V_0/[1+(x/x_0)^2]$ with energy below, slightly above and above
$V_0$, respectively [$V_0=0.03$ (arb. units) and $x_0=200$ (arb. units)]. 
e), f) and g) show the corresponding FT. Curves c), d), f) and g) are
shifted vertically for clarity.
The arrows indicate the range of wave vectors with significant weight
in the FT and, thus, confirm the measured behavior in Figs. 1g), 1h) 
and 1i).}

\end{figure*}

We exclude that a surface-projected bulk band, instead of a two
dimensional state, is responsible of this broad distribution of k
values in the 2D FT maps. Projected bulk bands' interference
patterns might appear only at close distances to the sample and with
oscillations corresponding only to wave vectors at the band edge
\cite{pascualprl}. Instead, the broad distribution of parallel
momentum has its origin in the spatial variation of the electric
field along the surface due to the finite curvature of the tip. The
local shift of the surface potential induced by the STM tip vanishes
gradually with the distance away from the tip position (Fig. 2a).
Accordingly, for a given electron energy E, the kinetic component
along the surface directions (E$_k$=E-E$_{FER}$, where E$_{FER}$ is
the FER binding energy)  increases continuously as the  electron is
accelerated away from the tip. Interference patterns carry
information of such inhomogeneous potential by showing oscillations
with shorter wavelength as the tip moves away from the scattering
potential, and a FT map with non-constant wave-vector distribution.
Deviations from a perfect spherical shape of the tip apex structure 
are probably responsible for the elliptical shape of the contours.

The evolution of dI/dV maps of Figs. 1d-1f and their FT images show
some interesting behavior.   First,  Fig. 1d  shows clear dI/dV
oscillations with relatively short wavelength although the image is
taken with eV$_s<E_{FER}$, i.e. below the FER onset (as shall be
shown later in subsection \ref{subsection2}, the peak position in
constant current dI/dV spectra like in Fig. 1c fits closely with the
energy E$_{FER}$). Second, for eV$_s$ values above E$_{FER}$ the FT
maps show an internal circle with a radius k$_{min}$ increasing with
the energy (Figs. 1h and 1i).  The origin of these phenomena can be
understood  by plotting the electron states of a one dimensional
quantum box with a decaying potential as in Fig. 2a.  In Figs. 2b we
consider the case of an electronic state with energy lower than
E$_{FER}$. In this case, no allowed state exist locally under the
STM tip. Only at distances beyond the point where the energy
E=E$_{FER}$ traveling states can exist. In the STM data (Fig. 1d)
this translates into large dark circles at scattering points, and
the onset of standing waves patterns beyond a certain distance from
the defects. In reciprocal space (Fig. 2e) a continuous window of wave
vector values from k$_{||}(0)=0$ to k$_{||}(s)$ reflects the continuous
acceleration of the electron waves from the tip position $r_{||}=0$ to
the scattering point $r_{||}=s$ at the surface. Following the same 1D
model, we expect that for electron energy above the resonance onset
E$_{FER}$ (Figs. 2c and 2d) a minimum wave vector k$_{||}(0)=
2m_e^*/\hbar^2(E-E_{FER})^{1/2}$ appears in the 1D plots (Figs. 2f and
2g), corresponding to the internal circle in Figs. 1h and 1i.

Therefore, the width of the wave vector distribution will
reflect the spatial change of the local (tip-induced) potential
shift. Ideally, for a given value of the tip radius, the critical
angle determined by the exponential decay of the Fowler-Nordheim
transmittivity \cite{gadzuk}, permits to estimate the maximum
parallel component of the electric field at the tip and, therefore,
the change of parallel momentum, based on simple classical
trajectory considerations. Assuming a radius of curvature of the tip
R=10 nm and a tip-sample distance Z=15 $\AA$ we estimate a change in
parallel momentum of 2 $nm^{-1}$ at 5 V, in agreement with our
previous analysis shown in figures 1g)-i).
Results similar to those shown in Fig. 1 are observed for other FERs that
appear below 8 eV. However, no wave patterns are seen above this value,
indicating a significant decrease of the electron lifetime.

\subsection {Line shape of dI/dV spectra \label{subsection2}}

To understand the role of the surface electronic structure
in the dynamics of FERs, next we explore the information contained in dI/dV
spectra about the energy width of FERs by analyzing their line shape and
compare with lifetime estimations based on a phase coherence length
analysis of interference wave patterns \cite{note}.

\begin{figure*} [tbp]
\includegraphics*[width=0.5\linewidth, angle=-90] {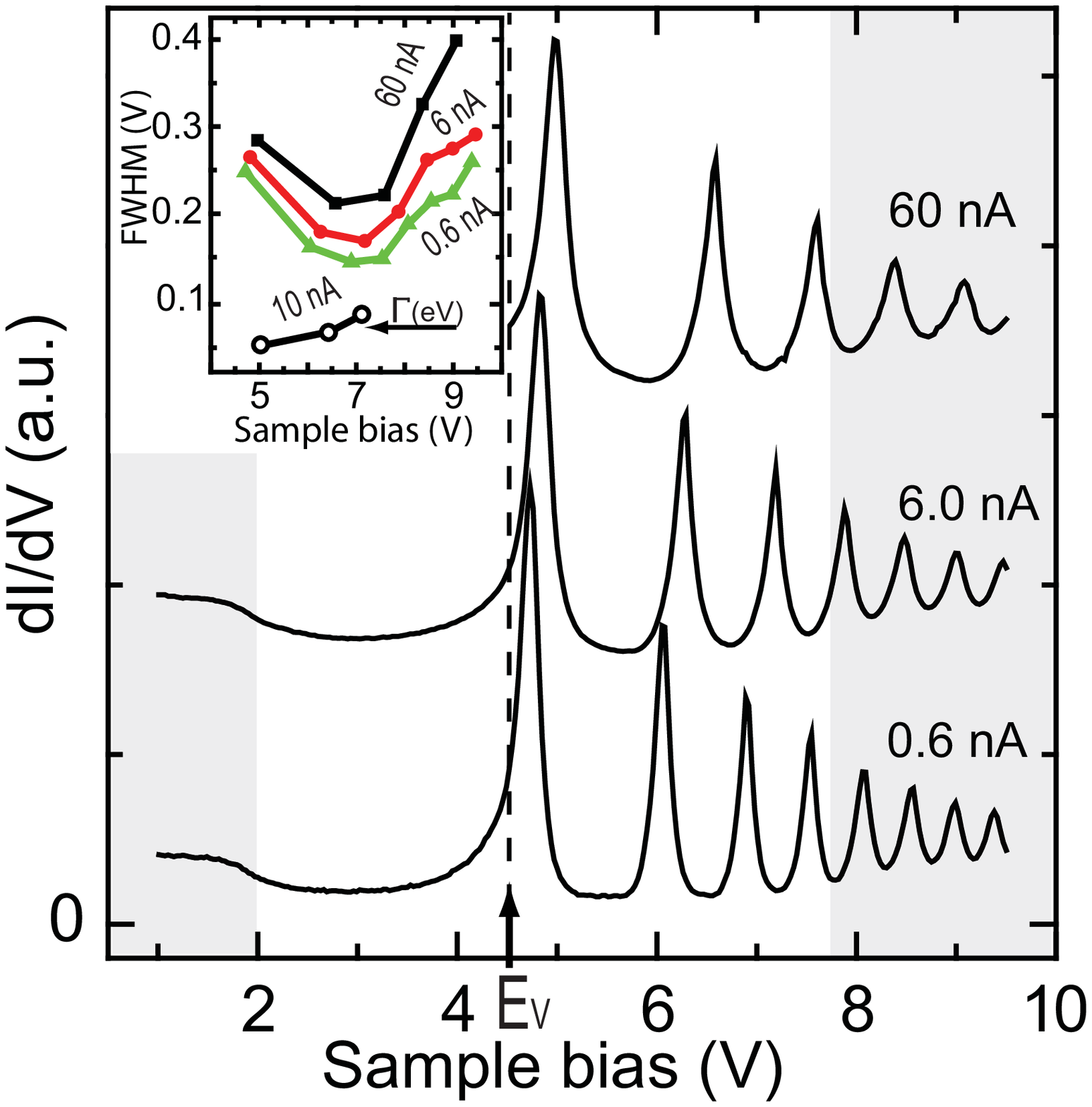}
\caption{(color online) dI/dV spectra taken for several current set points
ranging from 0.6 nA to 60 nA (V$_{ac}$=10 mV rms).
The inset shows the width of the peaks from a Lorentzian fit, as well as
values obtained from a phase coherence length analysis \cite{note}. 
Shaded grey areas correspond to the energy location in $\overline{\Gamma}$ 
of copper bulk bands projected on the (100) surface \cite{goldmann}.}
\end{figure*}

In Fig. 3 we show a series of constant current dI/dV spectra at
different set point current values. Sharp peaks appear in a wide
sample bias range  covering the energy range of the projected Cu
bulk gap and above \cite{goldmann}. Each peak corresponds to the
onset of a 2-D resonance state. In field emission regime, the tunnel
transmittivity T(E,V) is sharply peaked at the tip Fermi level
($E=E_{F}$); most of the current comes from a narrow energy window of
about 100 meV below $E_{F}$ \cite{gadzuk}, causing FERs to appear as
peaks in dI/dV spectra. On the contrary, in case of surface states 
close to $E_{F}$  dI/dV spectra show a line-shape close to a step
\cite{kliewer}. Interestingly, we find that the apparent width of
the dI/dV peaks exhibits a non-monotonic behavior (shown in the
inset). Resonances lying inside the gap are narrower than those
appearing above approximately 8 eV, whose width increases in
agreement with the steep decrease of surface reflectivity as the top
edge of the projected bulk band gap is crossed \cite{kasemo,goldmann}. 

The finite line-width of resonance states in the gap is expected to be 
dominated  by intrinsic factors like the electron reflectivity of the 
surface, spatial extension of the wave function \cite{echenique} and 
electric field strength at the tip-sample region \cite{crampin}. 
However, as it is shown in the inset of Fig. 3, the apparent width 
of the first three peaks in constant current dI/dV spectra is 
considerably larger than the intrinsic line-width estimates 
based on a phase coherence length analysis \cite{note}, which
for the first FER agree with previous \textit{ab initio} calculations 
\cite{crampin} and estimations \cite{kern} for the same system.
Therefore, an additional broadening mechanism must exist to explain the 
apparent width values in these dI/dV plots, which, presumably, is related 
with the method of measurement.

During the acquisition of \textit{dynamic} (constant
current) dI/dV spectra, the distance (Z) vs.  bias voltage (V)
characteristics (Z(V)) exhibit a pronounced step-like behavior as
the resonance is crossed for electron energies lying in the
projected gap. The resonant electron transmitivity in this close
feed-back loop spectroscopy is expected to be affected by the
continuous change of the tunneling barrier shape with bias voltage
and tip-sample distance. It is reasonable to assume that this
dynamic mode will introduce some  distortion in the resonance's line
shape respect the ideal {\em static} situation, in which the
tunneling barrier shape is kept fixed at Z(V) peak values. It is
only in this latter case that one could relate the width of peaks in
the transmitivity T(E) to the intrinsic width of resonances. We have
performed a model calculation  to establish a link between intrinsic
energy line-width of resonances in the {\em static} tunneling
transmitivity T(E) and the corresponding apparent width in constant
current dI/dV spectra, which is related to the  auxiliary {\em
dynamic} transmitivity T(E$_F$,V) at tip the Fermi level.

\begin{figure*} [tbp]
\includegraphics*[width=0.5\linewidth, angle=-90] {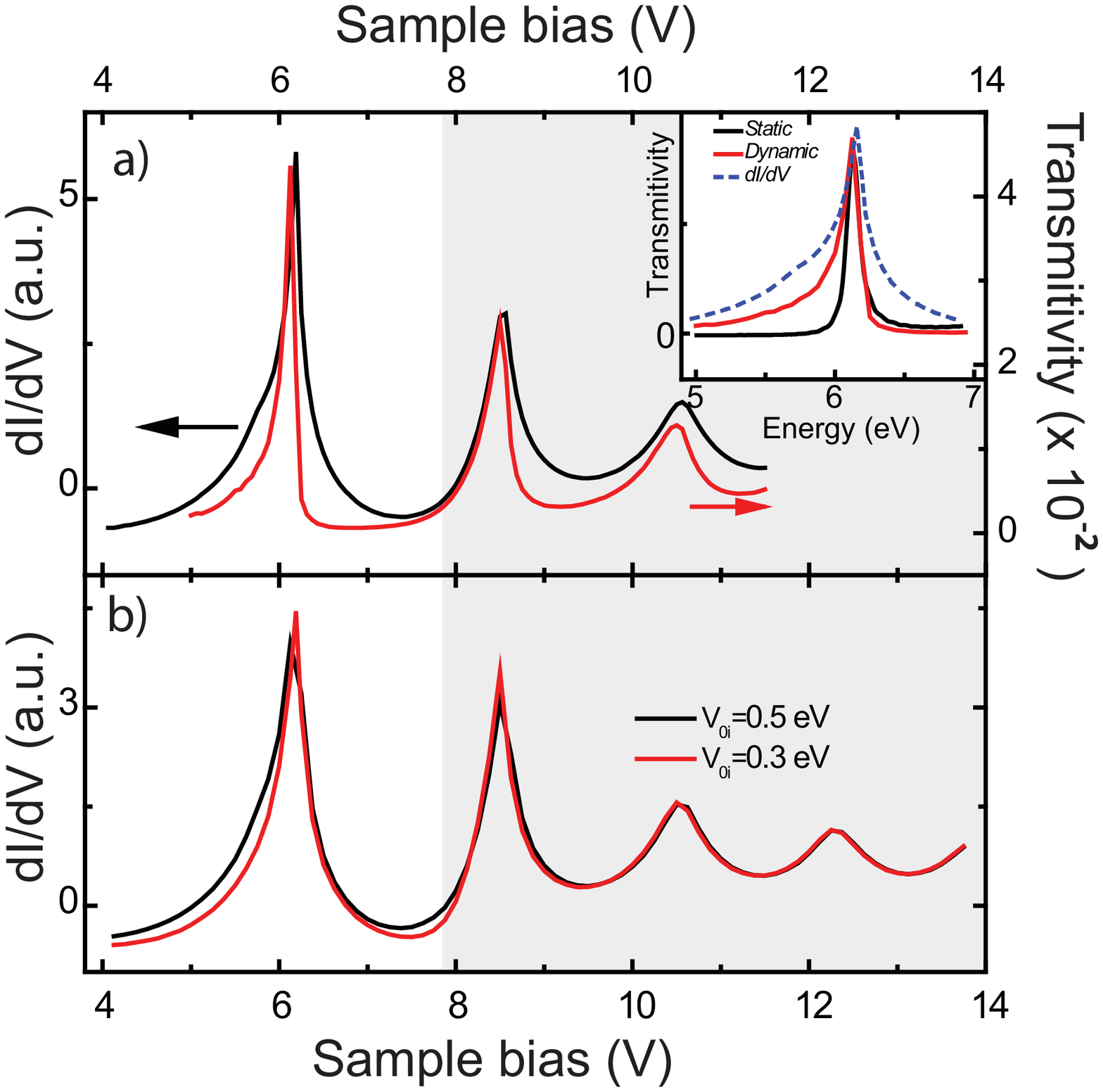}
\caption{(color online) a) Comparison between a calculated dI/dV spectrum
(black curve) and the transmission coefficient (red curve) at the tip Fermi
level T($E_{F}$;V). The inset shows a comparison between a {\it static}
calculation (black) of T(E) at Z(V) values of the first resonance
(thus reflecting the intrinsic FER line shape), T($E_{F}$;V) (red) and
dI/dV(V) (blue). b)dI/dV spectra for two $V_{0i}$ values (see the text).}
\end{figure*}

Our calculations are based on 1-D model potentials for the tip and
sample including the work function and Fermi energy that defines the
bottom of the surface potential. For the surface, a periodic
sinusoidal modulation that determines the magnitude and position of
the energy gap at the $\overline{\Gamma}$ point of the surface
Brillouin zone \cite{chulki} is included. Parallel wavevector
components are considered in the full 3-D calculations assuming a
free-electron-like (parabolic) dispersion.  To model electron
transmission inside the gap, inelastic scattering at the surface is
included by using a complex potential, similar to previous LEED
studies \cite{pendry,garcia}. Its imaginary part ($V_{0i}$)
introduces a decay of the electron flux due to absorption. A smooth
matching between the tip and sample potentials, which includes
multiple images, is used \cite{pitarke}.
The calculation of the tunneling current for a given
tip-sample distance (Z) and bias voltage requires the knowledge of the
energy dependence of the barrier transmission coefficient T(E,V) below
the tip's Fermi level $E_{F}$ \cite{duke}.
A quantitative agreement between measured and calculated
dI/dV spectra is not persecuted in this model approach, mostly  due
to the lack of knowledge of effective tunneling areas, tip-sample
distance or tip work function. Instead, we can provide a qualitative 
picture of the effect of the dynamic measurements on the peaks' width. 
The constant current dI/dV spectra are calculated by numerical 
differentiation of the current I(V,Z) along the constant current Z(V) 
characteristic.

In Fig. 4a we show a comparison between calculated dI/dV(V) spectra
and the corresponding dynamic transmitivity T($E_{F}$;V). Both
curves exhibit a similar shape and a characteristic increase of
their line-width with applied bias. This confirms the  high
collimation of field emitted electrons in a narrow energy window
below $E_{F}$. The inset compares the conductance dI/dV(V) around 
the first peak, the {\em dynamic} transmitivity T(E$_F$,V) and the 
intrinsic line-shape of the corresponding FER obtained in a 
{\it static} calculation at Z(V) peak values. Lorentzian fits to 
T(E$_F$,V) and dI/dV(V) give width values of $\sim$150 meV and 
$\sim$350 meV, respectively, while the intrinsic width of the
resonance in T(E) is $\sim$100 meV. Therefore, the broader line shape
in dI/dV spectra must be related both to the above mentioned finite energy
collimation and  the variation of the tunneling barrier shape with applied bias.
For FERs in the gap, the increase in electric field strength with bias shifts
the resonances to higher energy, appearing as broader peaks in
dI/dV(V). The increase of tip-sample distance in the dynamic method of
measurement partially compensates for this broadening effect.
The best conditions for a quantitative line-shape analysis can be
achieved at constant field strength conditions ($\sim$V/Z) and low set 
point current values (low applied field). 

By comparing (Fig. 4b) the shape of dI/dV curves calculated for two
different values of $V_{0i}$ we find that only the first peak in the
gap broadens as a response to the increase in absorption (inelastic
scattering). This confirms that inelastic scattering is the
broadening mechanisms of FERs lying inside the projected band gap.
At energies above the gap inelastic effects play a minor
role\cite{borisov} and the FER intrinsic line-width is dominated 
by elastic coupling to the bulk continuum (shaded area). In this region, 
the intrinsic resonances' line-width  is considerably larger 
(hence, no wave patterns in dI/dV maps could be seen for this energy)  
and, also, V/Z is almost constant. It is then expected that the effect 
of the dynamic broadening will be smaller, and the experimental dI/dV
peaks' apparent line-width will be close to the intrinsic value.

\section{CONCLUSIONS \label{section4}}

We find that the scattering of electrons in field emission states
with defects in a Cu(100) surface gives rise to isotropic standing
wave patterns, which reflect their dynamics in response to the
electric field gradients induced by the STM tip at the tip-sample
interface. Through a combined theoretical and experimental study
we have identified: (i) the origin of
characteristic peaks width in constant current dI/dV spectra above
the vacuum level as a combination of both, their FER intrinsic line
shape and the extrinsic distortion due to the measuring process, and
(ii) for FERs in the gap this distortion introduces an additional
broadening leading to a non-monotonic behavior of the width with
sample bias. Our results show that STS in the field emission regime 
can be used to gain information about the electron dynamics and 
surface electronic properties at energies well above the vacuum level.

\begin{acknowledgments}
We gratefully acknowledge Lucia Aballe, Andrei Borisov, Martin Hansmann,
Daniel S\'anchez Portal and Wolf Widdra for very helpful discussions. This
work was supported in part by the Basque Departamento de Educacion,
Universidades e Investigacion, the University of the Basque Country
UPV/EHU, the Spanish Ministerio de Educacion y Ciencia,  the EU Network of
Excellence NANOQUANTA (Grant No. NMP4-CT-2004-500198), and the EUROCORES
project MOL-VIC.
\end{acknowledgments}


\begin{thebibliography}{99}


\bibitem{nienhaus} H. Nienhaus, Surf. Sci. Rep. {\bf 45}, 1 (2002).

\bibitem{pedro} P. M. Echenique, R. Berndt, E.V. Chulkov, Th. Fauster,
A. Goldmann, and U. Höfer, Surf. Sci. Rep. {\bf 52}, 219 (2004).

\bibitem{bovensiepen} U. Bovensiepen, Prog. Surf. Sci. {\bf 78}, 87
(2005).

\bibitem{himpsel} F. J. Himpsel and J. E. Ortega, \prb {\bf 46}, 9719
(1992).

\bibitem{boger} K. Boger, M. Weinelt, and Th. Fauster, \prl {\bf 92},
126803 (2004).

\bibitem{kasemo} S. Andersson and B. Kasemo, Solid State Commun.
{\bf 8}, 961 (1970)

\bibitem{crommie} M. F. Crommie, C. P. Lutz, and D. M. Eigler, Nature
{\bf 363}, 524 (1993).

\bibitem{hasegawa} Y. Hasegawa and Ph. Avouris, \prl {\bf 71}, 1071
(1993).

\bibitem{hofmann} Ph. Hofmann, B. G. Briner, M. Doering,
H.-P. Rust, E. W. Plummer, and A. M. Bradshaw , \prl {\bf 79}, 265 (1997).

\bibitem{petersen} L. Petersen, P. T. Sprunger, Ph. Hofmann, E.
Lægsgaard, B. G. Briner, M. Doering, H.-P. Rust, A. M. Bradshaw, F.
Besenbacher, and E. W. Plumer, \prb {\bf 57}, R6858 (1998).

\bibitem{pascualprb01} J. I. Pascual, Z. Song, J. J. Jackiw, K. Horn,
and H.-P. Rust, \prb {\bf 63}, R241103 (2001).


\bibitem{pascualprl} J. I. Pascual, A. Dick, M. Hansmann, H.-P. Rust, J.
Neugebauer, and K. Horn, \prl {\bf 96}, 046801 (2006).


\bibitem{burgi} L. B\"urgi, O. Jeandupeux, H. Brune, and K. Kern,
\prl {\bf 82}, 4516 (1999).

\bibitem{kliewer} J. Kliewer, R. Berndt, E. V. Chulkov, V. M. Silkin, P.
M. Echenique and S. Crampin, , Science {\bf 288}, 1399 (2000).

\bibitem{kern} P. Wahl, M. A. Schneider, L. Diekhöner, R. Vogelgesang,
and K. Kern, \prl {\bf 91}, 106802 (2003).

\bibitem{pascualspin} J.I. Pascual,  G. Bihlmayer, Yu. M. Koroteev,
H.-P. Rust, G. Ceballos, M. Hansmann, K. Horn, E. V. Chulkov, S. Blügel,
P. M. Echenique, and Ph. Hofmann, \prl {\bf 93}, 196802  (2004).

\bibitem{corriol} C. Corriol, V. M. Silkin, D. Sánchez-Portal,
A. Arnau, E. V. Chulkov, P. M. Echenique, T. von Hofe,
J. Kliewer, J. Kröger, and R. Berndt, \prl {\bf 95}, 176802 (2005).

\bibitem{berndt_stark} L. Limot, T. Maroutian, P. Johansson,
and R. Berndt, \prl {\bf 91}, 196801 (2003).

\bibitem{binig} G. Binnig, K. H. Frank, H. Fuchs, N. Garcia, B. Reihl, H.
Rohrer, F. Salvan, and A. R. Williams, \prl {\bf 55}, 991 (1985).

\bibitem{becker} R. S. Becker, J. A. Golovchenko, and B. S.
Swartzentruber, \prl {\bf 55}, 987 (1985).

\bibitem{himpsel2} T. Jung, Y. M. Mo, and F. J. Himpsel, \prl {\bf 74},
1641 (1995).

\bibitem{schneider} M. Pivetta, François Patthey, Massimiliano
Stengel, Alfonso Baldereschi, and Wolf-Dieter Schneider,
\prb {\bf 72}, 115404 (2005).

\bibitem{kubby} J. A. Kubby, Y. R. Wang, and W. J. Green, \prl {\bf 65},
2165 (1990).

\bibitem{moleetal} A.J. Caama\~no, Y. Pogorelov, O. Custance, J. Mendez,
A.M. Baró, J.Y. Veuillen, J.M. Gómez-Rodríguez, J.J. Sáenz,
Surf. Sci. {\bf 426}, L420 (1999).

\bibitem{bobrov} K. Bobrov, A. J. Mayne and G. Dujardin, Nature {\bf 413},
616 (2001).

\bibitem{horn} H.-P. Rust, J. Buisset, E.K. Schweizer and L. Cramer,
Rev. Sci. Instrum. 68, 129 (1997).

\bibitem{gadzuk} J. W. Gadzuk, \prb {\bf 47}, 12832 (1993).

\bibitem{note} We can estimate a  phase coherence length following 
the analysis in Refs. \cite{burgi,kern} neglecting the effect of the
electric field on the wave patterns profile along the surface.  
It is implicitly assumed that when extracting the intrinsic 
lifetime $\tau=1/{\Gamma}$ from the measured phase coherence length 
$L_{\phi} = v_{\phi}\tau$, a mean value of $k_{||}$ exists such that 
$mv_{\phi}=hk_{||}$, where m is the free electron mass. 
The mean value of $k_{||}$ is extracted from the wave length of the real 
space wave patterns and $L_{\phi}$ from the exponential decay of the 
amplitude of the oscillations that appear close to a straight step 
shown in Fig. 1. Although this mean value of $k_{||}$ is not well 
defined, it would correspond to a value between the minimum and maximum
values shown as arrows in Figs. 2g. This "rough" estimation
is heuristically justified since the result obtained is similar to
the one obtained in Ref. \cite{kern} and since the wave vector
dependence of the electron lifetime  is not too  pronounced
\cite{crampin}.

\bibitem{goldmann} A. Goldmann, V. Dose and G. Borstel, \prb {\bf 32},
1971 (1985).

\bibitem{crampin} S. Crampin, \prl {\bf 95}, 046801 (2005).

\bibitem{echenique} P. M. Echenique and J. B. Pendry,
Progress in Surface Science {\bf 32}, 111 (1989).


\bibitem{chulki} E. V. Chulkov, V. M. Silkin, and P. M. Echenique,
Surf. Sci. {\bf 437}, 330 (1999).

\bibitem{pendry} J. B. Pendry, Low Energy Electron Difraction, (Academic
Press, London, 1974).

\bibitem{garcia} R. García, J. J. Sáenz, J. M. Soler and N. García
, Surf. Sci. {\bf 181}, 69 (1986).

\bibitem{pitarke} J.M. Pitarke, F. Flores, and P.M. Echenique, Surf.
Sci. {\bf 234}, 1 (1990).

\bibitem{duke} C. B. Duke, in "Tunneling in Solids" edited by F.
Seitz, D. Turnbull and H. Ehrenreich (Academic Press, New York, 1969).

\bibitem{borisov} A. Borisov, E. V. Chulkov, and P. M. Echenique,
\prb {\bf 73}, 073402 (2006).


\end{thebibliography}
\end{document}